%% file: ms.tex
\documentclass[a4paper,10pt]{article}
\pdfoutput=1

\usepackage{geometry}
\usepackage{subfig}
\usepackage{color}
\usepackage{amsmath}
\usepackage{amsfonts}
\usepackage{amssymb}
\usepackage{amsthm}
\usepackage{pgfplots}
\usepackage{tikz}
\usepackage{caption}
\usepackage{float}
\usepackage[parfill]{parskip}
\usepackage[hang,flushmargin]{footmisc}
\usepackage{verbatim}
\usepackage{bbm}
\usepackage{bm}
\usepackage{physics}
\usepackage{cite}
\usepackage{authblk}
\usetikzlibrary{calc}

\newcommand{\F}{\hat{F}}
\numberwithin{equation}{section}
\numberwithin{figure}{section}

\usepackage{pgffor}

\renewcommand*{\S}{%
  S
}
\newcommand*{\Sfactor}{%
  \tilde{S}
}
\newcommand*{\s}{%
 s
}
\renewcommand*{\F}{%
 F
}
\newcommand*{\Ffactor}{%
 \tilde{F}
}
\newcommand*{\f}{%
 f
}

\title{An exact power series representation of the Baker-Campbell-Hausdorff formula}

\author{J.C. Moodie\thanks{jxm276@student.bham.ac.uk} }
\author{M.W. Long}
\affil{
School of Physics and Astronomy, University of Birmingham, Edgbaston, Birmingham, B15~2TT,
United Kingdom.
}
\date{}

\begin{document}
\maketitle
\input{abstract}

\input{introduction.tex}
\input{mainResult.tex}

\input{finiteExamples.tex}

\input{choiceOfBasis.tex}
\input{conclusion.tex}

\appendix
\input{calculationOfTheSums.tex}

\bibliography{bib}{}

\end{document}

%% file: abstract.tex
\begin{abstract}
  An exact representation of the Baker-Campbell-Hausdorff formula as a power series in just one of the two variables is constructed.
  Closed form coefficients of this series are found in terms of hyperbolic functions, which contain all of the dependence of the second variable.
  It is argued that this exact series may then be truncated and expected to give a good approximation to the full expansion if only the perturbative variable is small.
  This improves upon existing formulae, which require both to be small.
  As such this may allow access to larger phase spaces in physical problems which employ the Baker-Campbell-Hausdorff formula, along with enabling new problems to be tackled.
\end{abstract}

%% file: introduction.tex
\section{Introduction}

In physics and mathematics\cite{Vajna2018, Van-Brunt2016, Achilles2012} it is often useful to write the product \( e^X e^Y \) as \( e^Z \), for some \( Z \).
When the objects \( X \) and \( Y \) do not commute, as is often the case when dealing with matrices, it may not be elementary to find such a \( Z \).
Many authors\cite{Achilles2012, Campbelll896, Campbell1897,Poincare1899a,Poincare1899b,Baker1901,Baker1902,Baker1905, Hausdorff1906} attempted to deal with this problem by targeting \( Z(X,\,Y) \equiv \log \left( e^X e^Y \right) \).
Such attempts resulted in the Baker-Campbell-Hausdorff formula, 
\[
  Z(X,\, Y) \equiv \log(e^X e^Y) = X + Y + \frac{1}{2} [X,\, Y] + \frac{1}{12} \Big( [X,\, [X,\, Y]] + [Y,\, [Y,\, X]] \Big) + \dots \,.
\]
Dynkin\cite{Dynkin1947} found this formula explicitly in terms of commutators for every order, where order means combined powers of \( X \) and \( Y \).
Unfortunately, this means that if a truncation of the series is to give a good approximation to the full expansion, both \( X \) and \( Y \) must be sufficiently close to zero.

There exists an alternative representation to all orders in \( X \) but linear in \( Y \).
Letting \( L_X Y \equiv [X,\, Y] \) denote commutator operators, it is given by
\begin{equation}\label{eq:regularBCH}
  Z(X,\, Y) = X + \frac{L_X}{2 \sinh \left( \frac{1}{2} L_X \right)} e^{\frac{1}{2} L_X} Y + \mathcal{O}(Y^2) \,.
\end{equation}
The aim of this work will be to extend this representation to all powers of \( Y \).
That is, find explicit expressions for \( \hat{G}_n \) given from
\[
  Z = X + \sum_{n=1}^\infty \hat{G}_n \left( e^{\frac{1}{2}L_X} Y \right)^n \,. 
\]
These operators \( \hat{G}_n \) will depend non-trivially on commutator operators \( L_X \).
This series may be truncated and give a good approximation to the full expansion if only \( Y \) is small, as opposed to both \( X \) and \( Y \) in the previous.

%% file: mainResult.tex
\section{Main result}
Consider a symmetric version of of the Baker-Campbell-Hausdorff formula,
\begin{equation} \label{eq:symmetricBCH}
  \mathcal{S}(A,\,B) \equiv \log \left( e^{A} e^{2B} e^{A} \right) \,,
\end{equation}
for two matrices \( A \) and \( B \).
While this formulation is more natural to work with than \eqref{eq:regularBCH}, each may be transformed into the other via the Hadamard formula.
Employing the notation for commutators which shall be used throughout this article, \( LB \equiv [A,\,B] \) and \( L^nB \equiv [A,\,[A,\,\dots,[A,\, B]\,\dots\,]\,] \), it states that
\begin{equation}\label{eq:Hadamard}
  e^A B e^{-A} = e^{L} B \,.
\end{equation}
This then implies
\[
  e^A e^B e^{-A} = e^{e^{L} B} \,,
\]
from which it is easily seen that \( \mathcal{S}(A,\, B) = Z(2A,\, 2 \exp(-L) B) = Z(2 \exp(L)B, 2A) \) and additionally \( Z(X,\,Y) = \mathcal{S}( X/2, \exp(L_X/2) Y/2) \).
The factors of two have been introduced here in order to simplify the final answer.

The task ahead is to expand equation \eqref{eq:symmetricBCH}.
The matrix \( B \) will be the focus, with the aim being to write the expansion as a power series in this matrix.
Once this is achieved, the coefficients of the power series will be examined in depth and closed form expressions obtained.

The identity
\begin{equation}\label{eq:identity}
  \log M = - \sum_{l=1}^\infty \frac{1}{l} \sum_{m=0}^l (-1)^m \frac{l!}{m! (l-m)!} M^m \,,
\end{equation}
may be employed, setting \( M = \exp(A) \exp(2B) \exp(A) \).
It will be found that \( M^m \) separates into several parts, each of which takes the form \( f \exp (2 m A) g \) for \( m \)-independent quantities \( f \) and \( g \).
Both \( f \) and \( g \) may then be pulled out of the above sums, and hence the logarithm, leaving \( \exp(2 m A) \) in place of \( M^m \).
The identity then may be used in reverse to obtain \( \log(M) = f\, 2A \, g \).

To that end, the focus will now be on calculating \( M^m \).
The Hadamard formula \eqref{eq:Hadamard} may be used to symmetrically move exponentials of \( A \) to the edges, obtaining
\[
  M^m = e^{m A} \left[ \prod_{n=-\frac{m-1}{2}}^{\frac{m-1}{2}} \exp\left(2 e^{2 n L} B\right) \right] e^{m A} \,,
\]
where the product must be taken in the correct order, namely increasing \( n \).
The exponentials involving \( B \) may then be Taylor expanded
\[
  M^m = e^{m A} \left[ \prod_{n=-\frac{m-1}{2}}^{\frac{m-1}{2}} \sum_{k_n=0}^\infty \frac{1}{k_n!} \left(2 e^{2 n L} B\right)^{k_n} \right] e^{m A} \,,
\]
and terms gathered in orders of \( B \),
\begin{multline*}
  M^m = e^{m A} \left[ 
    1 
    + 2\left( \sum_{-\frac{m-1}{2} \leq n_1 \leq \frac{m-1}{2}} e^{2 n_1 L} B \right) \right.\\ 
    \left.+ 2^2 \left( \sum_{-\frac{m-1}{2} \leq n_1 < n_2 \leq \frac{m-1}{2} }  e^{2 n_1 L} B e^{2 n_2 L} B + \frac{1}{2!}\sum_{-\frac{m-1}{2} \leq n_1 \leq \frac{m-1}{2}} e^{2 n_1 L} B e^{2 n_1 L} B \right)
    + \cdots \right] e^{m A}
\end{multline*}

In the above expression, each term \( \exp( 2 n_i L ) B \) must be thought of as one object - that particular commutator operator \( L \) is acting on that particular matrix \( B \) and so the two are intrinsically linked.
It is helpful to formalise this link, labelling the pair with an index.
Then it is understood that the operator \( L_i \) acts on only the matrix \( B_i \), and no other.
Each such pair may then be labelled.
This allows the commutation of operators and matrices with different labels, enabling all matrices \( B \) in the above expression to be pulled out of each sum.
Explicitly,
\begin{align}
  M^m &= e^{m A} \left[ 
    1 
    + 2 \left( \sum_{-\frac{m-1}{2} \leq n_1 \leq \frac{m-1}{2}} e^{2 n_1 L_1} \right)  B_1  \right. \notag\\
    &\left.\qquad\qquad+ 2^2 \left( \sum_{-\frac{m-1}{2} \leq n_1 < n_2 \leq \frac{m-1}{2}}  e^{2 n_1 L_1} e^{2 n_2 L_2}  + \frac{1}{2!}\sum_{-\frac{m-1}{2} \leq n_1 \leq \frac{m-1}{2}} e^{2 n_1 \left(L_1 + L_2 \right)} \right) B_1 B_2
    + \cdots \right] e^{m A} \label{eq:powerOfM} \\
    &\equiv e^{m A} \left[ 
    \F_0
    + \F_1(L_1)\, B_1
    + \F_2(L_1,\, L_2)\, B_1 B_2
    + \F_3(L_1,\, L_2,\, L_3)\, B_1 B_2 B_3
    + \cdots \right] e^{m A} \,. \label{eq:powerOfMCompact}
\end{align}

The first aim has thus been achieved; the formula \eqref{eq:symmetricBCH} has been expanded with a power series in the matrix \( B \).
The next is to find closed form expressions for the coefficients \( F_N \).
First define the sum \( \S_N \) as
\begin{equation}\label{eq:explicitSum}
  \S_N(L_1,\, L_2,\, \dots,\, L_N) \equiv 2^N \sum_{-\frac{m-1}{2} \leq n_1 < n_2 < \cdots < n_N \leq \frac{m-1}{2}} e^{2 n_1 L_1}  e^{2 n_2 L_2} \cdots  e^{2 n_N L_N} \,,
\end{equation}
then the first few of the coefficients \( \F_N \) are given by
\begin{gather}
  \F_0 = 1 \,,\notag\\
  \F_1(L_1) = \S_1(L_1) \,,\notag\\
  \F_2(L_1,\, L_2) = \S_2(L_1,\, L_2) + \frac{2}{2!} \S_1(L_1 + L_2) \,,\notag\\
  \F_3(L_1,\, L_2,\, L_3) = \S_3(L_1,\, L_2,\, L_3) + \frac{2}{2!} \Big( \S_2(L_1 + L_2,\, L_3) + \S_2(L_1,\, L_2 + L_3) \Big) + \frac{2^3}{3!} \S_1(L_1 + L_2 + L_3) \,. \label{eq:F3}
\end{gather}
Writing the coefficients \( \F_N \) for an arbitrary order \( N \) is a problem in partitioning.
As seen in the above examples, the string \( L_1 + L_2 + \cdots + L_N \) is split in all possible ways.
The resultant substrings are then used as arguments for the sums \( \S_n \).
However, each sum is also divided by factorials.
These factorials are determined by the length of the substrings used as arguments.
For example, the string \( L_1 + L_2 + L_3 \) may be split in the following ways giving the following factorials:
\begin{equation} \label{eq:recipe}
\begin{gathered}
 L_1 + L_2 + L_3 \quad\longrightarrow\quad \enspace \enspace3! \,, \enspace\enspace \\
 L_1 + L_2 \enspace,\; L_3 \quad\longrightarrow\quad \enspace 2!\,1!\,, \enspace  \\
 L_1 \enspace,\; L_2 + L_3 \quad\longrightarrow\quad \enspace 1!\,2! \,, \enspace \\
 L_1 \enspace,\; L_2 \enspace,\; L_3 \quad\longrightarrow\quad 1!\,1!\,1! \,,
\end{gathered}
\end{equation}
demonstrating how \( F_3 \) was constructed in equation \eqref{eq:F3}.

There are then two major hurdles to finding closed form expressions for each coefficient of the power series.
The first is to calculate the explicit sum \( S_N \).
As the sum \( S_N \) may be thought of as \( N \) finite geometric series, it may be expected to have \( 2^N \) terms.
However, it may be split into \( N+1 \) parts, each of which is a collection of infinite geometric series.
This lifting of the constraint is crucial and will be discussed shortly.
The second hurdle is then to perform the partition sum, that is to calculate \( F_N \) given the functions \( S_r \).

It is useful at this point to deal with a concrete example.
Consider the sum
\[
  S_2(L_1,\, L_2 ) \equiv \sum_{-\frac{m-1}{2} \leq n_1 < n_2 \leq \frac{m-1}{2} } 2^2 e^{n_1 L_1} e^{n_2 L_2} \,.
\]
The summation variables, \( n_1 \) and \( n_2 \), are constrained from above and below.
These constraints may be thought of as forming a triangle, as depicted in Figure \ref{fig:S2}.
The sum may then be thought of as the combination of three semi-constrained sums, constructed by taking a given vertex of the triangle and extending the constraining lines to form infinite triangles or rectangles.
Explicitly,
\begin{multline}\label{eq:triangles}
\sum_{-\frac{m-1}{2} \leq n_1 < n_2 \leq \frac{m-1}{2} } 2^2 e^{2 n_1 L_1} e^{2 n_2 L_2} = 
 2^2 \left( \sum_{n_1 < n_2 \leq \frac{m-1}{2}} e^{2 n_1 L_1} e^{2 n_2 L_2} \right)
  - 2^2 \left( \sum_{n_1 < -\frac{m-1}{2}} e^{2 n_1 L_1} \sum_{n_2 \leq \frac{m-1}{2}} e^{2 n_2 L_2} \right) \\
  + 2^2 \left( \sum_{n_2 \leq n_1 < -\frac{m-1}{2}} e^{2 n_1 L_1} e^{2 n_2 L_2}  \right)\,.
\end{multline}
The sums on the right-hand side may then be evaluated to obtain
\[
  S_2 (L_1,\, L_2 ) = \frac{ \coth( L_1 ) - 1 }{\sinh(L_1 + L_2)} e^{m (L_1 + L_2)} + \frac{1}{\sinh(-L_1)}\frac{1}{\sinh(L_2)} e^{m (-L_1 + L_2)} + \frac{ \coth( - L_2 ) - 1 }{\sinh(-L_1 - L_2)} e^{m (-L_1 - L_2)}  \,.
\]
\input{S2.tex}

Generalising this idea to the sum \( S_N \) involves \( N + 1 \) vertices of an \(N\)-dimensional tetrahedron.
The constraining lines are extended, creating \( N + 1 \) sums similar to those in equation \eqref{eq:triangles}.
This is done carefully in appendix \ref{app:calculationOfTheSums}, with the final result
\begin{multline} \label{eq:Sresult}
  \S_N(L_1,\,\dots,\, L_N) = \sum_{r=0}^N \Sfactor_r(-L_r, - L_{r-1} - L_r\,, \dots,\, - L_1 - L_2 - \cdots- L_r)\\
  \times \Sfactor_{N-r}(L_{r+1},\, L_{r+1} + L_{r+2},\, \dots,\,L_{r+1} + L_{r+2} + \cdots + L_N) e^{m(-L_1 - \cdots - L_r + L_{r+1} + \cdots + L_N)} \,,
\end{multline}
where 
\begin{equation}\label{eq:Sfactor}
  \Sfactor_r(x_1,\, \dots,\, x_r) = \frac{\s_{r-1}(x_1,\, \dots,\, x_{r-1})}{\sinh\left(x_r \right)} \qquad \text{for } r \in \mathbb{Z}^+\,, \qquad \Sfactor_0 \equiv 1 \,,
\end{equation}
and
\begin{equation}\label{eq:s}
  \s_{r-1}(x_1,\, \dots,\, x_{r-1}) = \prod_{j=1}^{r-1} \left[ \coth( x_j ) - 1 \right]  \qquad \text{for } (r-1) \in \mathbb{Z}^+\,, \qquad \s_0 \equiv 1 \,.
\end{equation}
There are several things to note from this result.
Firstly, the \( N + 1 \) different forms that the exponential above may take clearly correspond to the vertices of the \( N \)-dimensional tetrahedron discussed previously.
As mentioned earlier, this exponential, containing all the \(m\)-dependence, is crucial in reversing the identity \eqref{eq:identity}.
Next, note the splitting of each term into \( \Sfactor_r \) and \( \Sfactor_{N-r} \) functions.
This structure remains for the coefficients \( \F_N \), as shall be seen shortly, and appears fundamental to the problem.
Furthermore, the representation of the result in hyperbolic functions is perhaps not unexpected; previous results showed that the order \( B \) term is best written with a \( \sinh \) function.
Finally, the arguments of these factorised functions at first appear rather cumbersome.
However, the constituent hyperbolic functions only ever contain sums of the commutator operators \( L_i \).
As such it is mathematically sensible to think of the active variables not as these commutator operators \( L_1,\, L_2,\, L_3 \) etc., but rather as strings of such operators, for example \(L_1,\, L_1 + L_2,\, L_1 + L_2 + L_3 \) etc.
More will be said of such strings in later sections \ref{sec:choiceOfBasis}.

The next task is to perform the partition sum, or in other words calculate \( F_N \) given the now known \( S_r \).
Once again it is useful to turn to an example.
Using the above results, it is simple to read off that
\begin{align*}
  \F_3(L_1,\, L_2,\, L_3) &\equiv \S_3(L_1,\, L_2,\, L_3) + \frac{2}{2!} \Big[ \S_2(L_1,\, L_2 + L_3) + \S_2(L_1 + L_2,\, L_3) \Big] + \frac{2^2}{3!} \S_1(L_1 + L_2 + L_3) \\
  &= C_0 e^{m(L_1 + L_2 + L_3)} + C_1 e^{m(-L_1 + L_2 + L_3)} + C_2 e^{m(-L_1 - L_2 + L_3)} + C_3 e^{m(-L_1 - L_2 - L_3)}
\end{align*}
where
\begin{align*}
  C_0 &= 
    \frac{
      \big( \coth ( L_1 ) - 1 \big) \big( \coth ( L_1 + L_2 ) - 1 \big) 
      + \dfrac{2}{2!} \Big[ \big( \coth ( L_1 ) - 1 \big) + \big( \coth ( L_1 + L_2 ) - 1 \big) \Big]
      + \dfrac{2^2}{3!}
    }{\sinh \left( L_1 + L_2 + L_3 \right)}
  \,, \\
  C_1 &= 
    \left[ \vphantom{\frac{\dfrac{2^2}{2!}}{\sinh(L_1)}} 
      \frac{1}{\sinh \left(- L_1 \right)} 
    \right]
    \left[ 
      \frac{
        \big( \coth ( L_2 ) - 1 \big) 
        + \dfrac{2}{2!}
      }{\sinh \left(L_2 + L_3 \right)} 
    \right]
  \,,\\
  C_2 &= 
    \left[ 
      \frac{
        \big( \coth ( -L_2 ) - 1 \big) 
        + \dfrac{2}{2!}
      }{\sinh \left(- L_1 -L_2 \right)}  
    \right]
    \left[ \vphantom{\frac{\dfrac{2^2}{2!}}{\sinh(L_1)}}  
      \frac{1}{\sinh \left( L_3 \right)} 
    \right]
  \,, \\
  C_3 &= 
    \frac{
      \big( \coth ( -L_3 ) - 1 \big) \big( \coth ( -L_2 - L_3 ) - 1 \big) 
      + \dfrac{2}{2!} \Big[ \big( \coth ( -L_3 ) - 1 \big) + \big( \coth ( -L_2 - L_3 ) - 1 \big) \Big]
      + \dfrac{2^2}{3!}
    }{\sinh \left( -L_1 - L_2 - L_3 \right)}
  \,.
\end{align*}
Within this example many of the previous themes are exposed.
Recall that the structure of the sum \( S_N \) had three major components: the result split into \( N + 1 \) terms; each term separated into an \(m\)-dependent exponential and an \(m\)-independent function; and each function then could be factorised.
It is clear that the first two of these must remain true for the partition sum \( F_N \), as there are \( N + 1 \) different forms the exponential may take.
Next, the sums which contribute to the coefficient of \( \exp( m (-L_1 - \cdots - L_r + L_{r+1} + \cdots + L_N) ) \) must contain a partition between \( L_r \) and \( L_{r+1} \).
Any other partitioning which occurs to the left of the split affects a given sums contribution to the term independently of any partitioning to the right.
As such, the factorisation must also remain true.
Combining these observations,
\begin{multline}\label{eq:Fresult}
  F_N(L_1,\, L_2,\, \dots,\, L_N ) = \sum_{r=0}^N \Ffactor_r(-L_r, - L_r - L_{r-1}\,, \dots,\,- L_r - L_{r-1} - \cdots - L_1)\\
  \times \Ffactor_{N-r}(L_{r+1},\, L_{r+1} + L_{r+2},\, \dots,\,L_{r+1} + L_{r+2} + \cdots + L_N) e^{m(-L_1 - \cdots - L_r + L_{r+1} + \cdots + L_N)} \,,
\end{multline}
where 
\begin{equation}\label{eq:Ffactor}
  \Ffactor_r(x_1,\, \dots,\, x_r) = \frac{\f_{r-1}(x_1,\, \dots,\, x_{r-1})}{\sinh\left(x_r \right)} \qquad \text{for } r \in \mathbb{Z}^+\,, \qquad \Ffactor_0 \equiv 1 \,.
\end{equation}
The function \( \f_{r-1}(x_1,\, x_2,\,\dots,\, x_{r-1}) \) will be a partition sum of the functions \( \s_n \) given in \eqref{eq:s}.
For example, the function
\begin{align*}
  \f_2(x_1,\,x_2) &= \s_2(x_1,\,x_2) + \frac{2}{2!} \Big[ s_1(x_1) + s_1(x_2) \Big] + \frac{2^2}{3!} \\
  &= \big( \coth ( x_1 ) - 1 \big) \big( \coth ( x_2 ) - 1 \big) 
      + \frac{2}{2!} \Big[ \big( \coth ( x_1 ) - 1 \big) + \big( \coth ( x_2 ) - 1 \big) \Big]
      + \frac{2^2}{3!} \,,
\end{align*}
is found in both \( C_0 \) and \( C_3 \) above.
In general, \( \f_{r-1}(x_1,\, x_2,\, \dots,\, x_{r-1}) \) is a sum of terms, each involving a product of \( \coth \) functions.
As shown for \( \f_2 \), in some of these terms there will be \( \coth \) functions missed out, 
In a term where \( n \) such functions in a row have been missed out, the coefficient will be \( a_{n+1} \equiv 2^{n}/(n+1)! \).
This then implies
\begin{equation}\label{eq:c-1}
  \f_{r-1} = \sum_{n=0}^{r-1} \sum_{\mathbf{p}} a_{p_0} a_{p_1} \cdots a_{p_n} \big( \coth(x_{p_0}) - 1 \big) \big( \coth(x_{p_0 + p_1}) - 1 \big) \cdots \big( \coth(x_{p_0 + \cdots + p_{n-1}}) - 1 \big)\,,
\end{equation}
where the sum over \( \mathbf{p} \equiv (p_0,\, p_1,\, \dots,\, p_n) \) is such that \( p_0 + \cdots + p_n = r \) and each \( p_i \in \mathbb{Z}^+ \).
While superficially complicated, this sum is actually very simple.
In a given term of the sum there are \( n \) \( \coth \) functions, which means \( (r-1)-n \) are missing.
However, it is important how they were missed out - if \( i \) in a row are missed out then from the reasoning above they must be replaced with the number \( a_i \).

Examining the above equation, it is clear that if \( p_0 > 1 \) then there have been \( p_0 - 1\) missed out prior to the first \( \coth \) and so a coefficient of \( a_{p_0} \) is picked up.
Similarly if \( p_1 > 1 \) then \( p_1 - 1 \) have been missed out between the first and the second and \( a_{p_1} \) is required.
Continuing this logic gives all terms in the above sum.

A simpler form of this function may be obtained.
The brackets in the sum may be expanded, putting the function into the form
\begin{equation}\label{eq:f}
  \f_{r-1} = \sum_{n=0}^{r-1} \sum_{\mathbf{p}} t_{p_0} t_{p_1} \cdots t_{p_n} \, \coth(x_{p_1})  \coth(x_{p_1 + p_2})  \cdots  \coth(x_{p_1 + \cdots + p_n}) \,.
\end{equation}
Comparing the constant term of \eqref{eq:c-1} with that of the above gives
\[
  t_{r} = \sum_{n=0}^{r-1} (-1)^{r-n+1} \sum_{\mathbf{p}} a_{p_0} a_{p_1} \cdots a_{p_n} \,.
\]
If this sum can be determined then this will of course give all numbers \( t_i \) which appear in equation \eqref{eq:f} by letting \( r = i \).
Generating functions may be employed to perform this calculation.
First multiply both sides by \( x^r \), and sum over \( r \):
\begin{align*}
  \sum_{r=0}^\infty t_{r} x^r
    &= \sum_{r=0}^\infty \sum_{n=0}^{r-1} (-1)^{n+1} \sum_{\mathbf{p}} a_{p_0} a_{p_1} \cdots a_{p_n} (-x)^r \\
    &= \sum_{n=0}^\infty \sum_{r=n+1}^\infty (-1)^{n+1} \sum_{\mathbf{p}} a_{p_0} a_{p_1} \cdots a_{p_n} (-x)^r \,.
\end{align*}
Next note
\[
  \sum_{r=n+1}^\infty (-1)^{n+1} \sum_{\mathbf{p}} a_{p_0} a_{p_1} \cdots a_{p_n} (-x)^r
    = \left[ -\sum_{k=1}^\infty a_k (-x)^k \right]^{n+1} \\
    = \left[ \frac{1 - e^{-2x}}{2} \right]^{n+1}
\]
and hence
\[
  \sum_{r=0}^\infty t_{r} x^r 
    = \sum_{n=0}^\infty \left[ \frac{1 - e^{-2x}}{2} \right]^{n+1} 
    = \tanh(x) \,,
\]
demonstrating the numbers \( t_r \) are generated by \( \tanh \).
When combined with equation \eqref{eq:f}, this then gives a clean formula for \( \f_{r-1} \) and thus \( \F_N \).
That is, \( f_{r-1} \) is a sum of products of \( \coth \) functions.
In each term of this sum, some number of these functions in a row will be missed out an replaced with the numbers \( t_r \) which come from the Taylor expansion of \( \tanh(x) \).
Finite examples of this concept will be given for clarity in section \ref{sec:finiteExamples}.

Focus will now turned to the exponential in equation \eqref{eq:Fresult}.
It is here that the identity \eqref{eq:identity} will be reversed.
Equation \eqref{eq:powerOfMCompact} may now be rewritten as
\[
  M^m = \sum_{N=0}^\infty e^{m A} \left( \sum_{r=0}^N \Ffactor_r \Ffactor_{N-r} e^{m (-L_1 - \cdots - L_r + L_{r+1} + \cdots + L_N)} \right) B_1 \cdots B_N e^{m A} \,,
\]
where the arguments of the functions have been suppressed for brevity.
Upon repeated application of the Hadamard formula \eqref{eq:Hadamard} this can be seen as
\begin{equation}\label{eq:preIdentity}
  M^m = \sum_{N=0}^\infty \sum_{r=0}^N \Ffactor_r \Ffactor_{N-r} \, B_1\cdots B_r \, e^{2mA} \, B_{r+1} \cdots B_N \,.
\end{equation}
The identity \eqref{eq:identity} may then be employed in reverse, obtaining
\[
  \log M = \sum_{N=0}^\infty \sum_{r=0}^N \Ffactor_r \Ffactor_{N-r} \, B_1\cdots B_r \, 2 A \, B_{r+1} \cdots B_N \,.
\]
Using the commutator operators \( L_i \), the matrix \( A \) in the above expression may be moved to either side of the matrices \( B \), via
\begin{align*}
B_1 \cdots B_r\, A \,B_{r+1} \cdots B_N 
  &= \left(- L_{1} - L_{2} - \cdots - L_{r}\right) B_1 \cdots B_N  +  A\, B_1 \cdots B_N \\
  &= \left(L_{r} + L_{r+1} + \cdots + L_{N}\right) B_1 \cdots B_N  +  B_1 \cdots B_N\, A \,.
\end{align*}
The case \( m = 0 \) in equation \eqref{eq:preIdentity} implies that for all \( N > 0 \),
\[
  \sum_{r=0}^N \Ffactor_r \Ffactor_{N-r} = 0 \,,
\]
which then enables \( \log M \) to be written in the form
\[
  \log M = \sum_{N=0}^\infty \left[ \sum_{r=0}^N \Ffactor_r \Ffactor_{N-r} \,(-L_1 - \cdots - L_r + L_{r+1} + \cdots + L_N ) \right] B_1 \cdots B_N \,.
\]
This then gives the promised expansion in powers of the matrix \( B \).
To summarise
\[
  \log\left(e^{A} e^{2B} e^A\right) = \sum_{N=0}^\infty  \hat{G}_N B_1 \cdots B_N \,, 
\]
where 
\begin{multline}\label{eq:FresultSummary}
 \hat{G}_N  = \sum_{r=0}^N \Ffactor_r(-L_r, - L_r - L_{r-1}\,, \dots,\,- L_r - L_{r-1} - \cdots - L_1)\\
  \times \Ffactor_{N-r}(L_{r+1},\, L_{r+1} + L_{r+2},\, \dots,\,L_{r+1} + L_{r+2} + \cdots + L_N)\\
  \times (-L_1 - \cdots - L_r + L_{r+1} + \cdots + L_N ) \,,
\end{multline}
\begin{equation}\label{eq:FfactorSummary}
  \Ffactor_r(x_1,\, \dots,\, x_r) = \frac{\f_{r-1}(x_1,\, \dots,\, x_{r-1})}{\sinh\left(x_r \right)} \qquad \text{for } r \in \mathbb{Z}^+\,, \qquad \Ffactor_0 \equiv 1 \,,
\end{equation}
and
\begin{equation}\label{eq:fSummary}
  \f_{r-1} = \sum_{n=0}^{r-1} \sum_{\mathbf{p}} t_{p_0} t_{p_1} \cdots t_{p_n} \, \coth(x_{p_1})  \coth(x_{p_1 + p_2})  \cdots  \coth(x_{p_1 + \cdots + p_n}) \,.
\end{equation}
Here the numbers \( t_k \) are given from the Taylor expansion of \( \tanh(x) \), and \( \mathbf{p} \equiv (p_0,\, p_1,\, \dots,\, p_n) \) is such that \( p_0 + \cdots + p_n = r \) and each \( p_i \in \mathbb{Z}^+ \).

%% file: S2.tex
\begin{figure}[h!]
\centering
\begin{tikzpicture}

  \draw[->, color=gray] (-5,0) -- (3,0);
  \draw[->, color=gray] (0,-5) -- (0,3);
  \node at (3, 0) [above, color=gray] {\(n_1\)};
  \node at (0, 3) [right, color=gray] {\(n_2\)};

  \draw[thick,dashed] (2,2) -- (-2,-2);
  \draw[thick] (2,2) -- (-5,2);
  \draw[thick] (-2,-2) -- (-5,-5);
  \draw[thick] (-1.98,2) -- (-1.98,-2);
  \draw[thick,dashed] (-2.02,2) -- (-2.02,-5);
  
  \node[above] at (2,2) {\( (\frac{m-1}{2},\,\frac{m-1}{2}) \)};
  \draw[fill=red] (2,2) circle (0.1);
  \node[above] at (-2,2) {\( (-\frac{m-1}{2},\,\frac{m-1}{2}) \)};
  \draw[fill=red] (-2,2) circle (0.1);
  \node[left] at (-2,-2)  {\( (-\frac{m-1}{2},\,-\frac{m-1}{2}) \)};
  \draw[fill=red] (-2,-2) circle (0.1);

\end{tikzpicture}
  \caption{
  A depiction of the parameter space of \( n_1 \) and \( n_2 \) in equation \eqref{eq:triangles}.
  Solid lines imply inclusiveness of that line in a given sum, while dashed imply the line of parameters is excluded.
  The variables of the original sum are constrained to the triangle formed from the vertices marked with a red circle.
  } \label{fig:S2}

\end{figure}
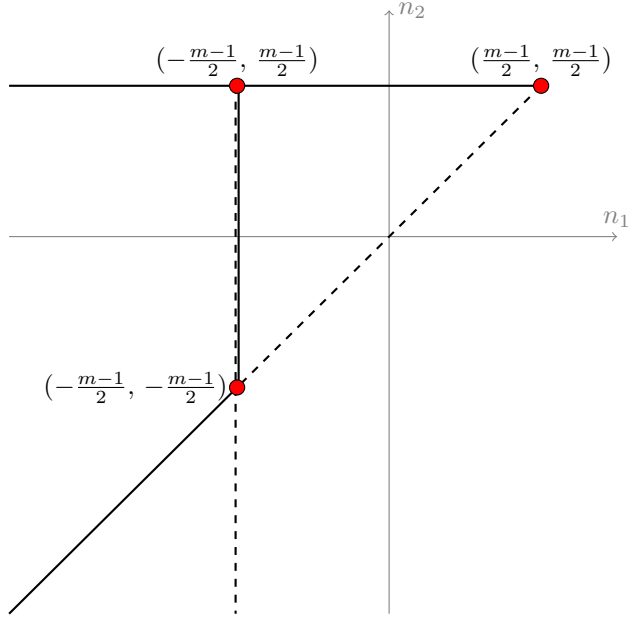

%% file: finiteExamples.tex
\section{Finite examples} \label{sec:finiteExamples}
While the general formula has been derived in the preceding section, it may be helpful to examine several low-order terms explicitly. 
This section will begin with the functions \( f_r \), for \( r = 0 \dots 5 \), highlighting the patterns previously discussed.
From these the operators \( \hat{G}_N \), the targets of this work, may be immediately written down and indeed will be for \( N = 1 \dots 3 \).

To begin, consider the functions \( f_{r} \).
The first few of these functions are given by
\begin{align*}
f_0 &\equiv 1 \,,\\
f_1 &= c_1 \,,\\
f_2 &= c_1 c_2 - \frac{1}{3} \,,\\
f_3 &= c_1 c_2 c_3 - \frac{1}{3} \, \big( c_1 + c_3 \big) \,,\\
f_4 &= c_1 c_2 c_3 c_4 - \frac{1}{3}\, \big( c_1 c_2 + c_1 c_4 + c_3 c_4\big) + \frac{2}{15} \,,\\
f_5 &= c_1 c_2 c_3 c_4 c_5 - \frac{1}{3} \, \big( c_1 c_2 c_3 + c_1 c_2 c_5 + c_1 c_4 c_5 + c_3 c_4 c_5\big) + \left(-\frac{1}{3}\right)^2  \, c_3 + \frac{2}{15}  \, \big( c_1 + c_5 \big) \,.\\
\end{align*}
Here the structure previously discussed becomes apparent.
In equation \eqref{eq:fSummary} the term in the sum where \( n = r-1 \)  forces each \( p_i \) to be equal to one, giving the full product of \( \coth \) functions with none missing.
This is the leading term in each of the examples above.
To generate the rest of the terms, neighbouring pairs of \( \coth \) functions in this term are replaced with \( - 1 / 3 \), neighbouring quadruplets are replaced with \( 2 / 15 \), and so on.
All possible such replacements appear in the above functions.

The targets of this work, the operators \( \hat{G}_N \), will now be examined.
It was previously mentioned that the leading term \( \hat{G}_1 \) is already well known and while this was calculated for the regular Baker-Campbell-Hausdorff formula \( Z(X,\,Y) \), it is of course trivial to map it to the symmetric version \( \mathcal{S}(A,\,B) \) considered here.
Using the general formulae of the preceding section, it would be natural to write
\[
  \hat{G}_1 = \left[ \frac{1}{\sinh ( L_1 )} \right] L_1 
      + \left[ \frac{ 1 }{\sinh(-L_1)} \right] ( -L_1 ) \,.
\]
Of course, as both \( x \) and \( \sinh(x) \) are odd functions, the minus signs are irrelevant and there is only really one term.

Next, at second order and third order it is found that
\[
  \hat{G}_2 = \left[ \frac{\coth (L_1)}{\sinh ( L_1 + L_2 )} \right]  ( L_1 + L_2 )  
      + \left[\frac{1}{\sinh(-L_1)} \right] \left[ \frac{1}{\sinh(L_2)}\right]  (-L_1 + L_2) 
      +  \left[\frac{\coth (-L_1)}{\sinh ( -L_1 - L_2 )} \right] ( -L_1 - L_2 ) \,,
\]
and
\begin{multline*}
  \hat{G}_3 = \left[ \frac{\coth (L_1) \coth (L_1 + L_2) - \frac{1}{3}}{\sinh ( L_1 + L_2 + L_3 )} \right] ( L_1 + L_2 + L_3)  
      + \left[ \frac{1}{\sinh(-L_1)} \right] \left[ \frac{\coth( L_2)}{\sinh(L_2 + L_3)} \right]  (-L_1 + L_2 + L_3)  \\
      + \left[ \frac{\coth( -L_2)}{\sinh(-L_1 - L_2)} \right] \left[ \frac{1}{\sinh(L_3)} \right]  (-L_1 - L_2 + L_3) 
      + \left[ \frac{\coth (-L_3) \coth (-L_2 - L_3) - \frac{1}{3}}{\sinh ( -L_1 - L_2 - L_3 )} \right] ( -L_1 - L_2 - L_3)  \,.
\end{multline*}
With these, some general themes begin to emerge.
It is immediately seen that each term factorises into two parts, written above with square brackets.
In a given term, all commutator operators with a plus sign gather into one of these parts while those with a minus sign gather into the other.
The only question that remains is how the arguments to each \( \coth \) function are determined. 

Consider, for example, the term involving \( - L_1 -L_2 - L_3 + L_4 + L_5 + L_6 + L_7 \) in \( \hat{G}_7 \).
Pictorially, the arguments for each function can be found from the diagram

\begin{figure}[h!]
  \centering
  \begin{tikzpicture}
    \node at (0,0) { \( -L_1-L_2-L_3+L_4+L_5+L_6+L_7 \, .\) };
    \draw[thick, red] (-.5,-.20) -- (-.5,-.30) -- (-2.8,-.30) -- (-2.8,-.20); 
    \draw[thick, blue] (-.5,-.35) -- (-.5,-.45) -- (-2.1,-.45) -- (-2.1,-.35); 
    \draw[thick, blue] (-.5,-.50) -- (-.5,-.60) -- (-1.2,-.60) -- (-1.2,-.50); 
    
    \draw[thick, red] (-.4,-.2) -- (-.4,-.30) -- (2.8,-.30) -- (2.8,-.2); 
    \draw[thick, blue] (-.4,-.35) -- (-.4,-.45) -- (2.,-.45) -- (2.,-.35); 
    \draw[thick, blue] (-.4,-.50) -- (-.4,-.60) -- (1.2,-.60) -- (1.2,-.50); 
    \draw[thick, blue] (-.4,-.65) -- (-.4,-.75) -- (.4,-.75) -- (.4,-.65); 
  \end{tikzpicture}
\end{figure}

Here, the top red lines highlight the arguments of each \( \sinh \) function, while the blue lines show the arguments to the \( \coth \) functions.
Combined with the previous discussion on how to write down these \( \coth \) functions to form the numerators, this says how to write \( \hat{G}_N \) for any order \( N \).
Of course equation \eqref{eq:FresultSummary} already provides such a formula, but perhaps observing these patterns for finite results may provide a more intuitive understanding.

%% file: choiceOfBasis.tex
\section{Choice of basis}\label{sec:choiceOfBasis}

In this section the sums of commutator operators, that is strings like \( L_1 + L_2 + \dots + L_r \), will be discussed.
It was previously suggested that these were mathematically natural to use as arguments to various functions.
It turns out that in the basis where the matrix \( A \) is diagonal, if such a basis exists, these sums result in the difference between two eigenvalues of \( A \).
As shall be seen, this drastically reduces the complexity of using the new representation.

First consider the matrix elements of \( L B \equiv [A,\, B] \):
\[
  \big[ L B \big]_{n_1 n_2} = A_{n_1 n'} B_{n' n_2} - B_{n_1 n'} A_{n' n_2} \,,
\]
where summation over repeated indices is assumed.
If \( A \) is a diagonal matrix then its matrix elements are given in terms of its eigenvalues as \( A_{n m} = a_n \delta_{nm} \), where \( \delta_{nm} \) is the Kronecker delta.
Hence in the basis \( A \) is diagonal the above is given by 
\[
  \big[ L B \big]_{n_1 n_2} = ( a_{n_1} - a_{n_2} ) B_{n_1 n_2} \,.
\]

More generally, for any Taylor expandable function \( g \), it can be seen that
\[
  \Big[ g(L_i + L_{i+1} + \cdots + L_{i+r}) B_1 B_2 \cdots B_N \Big]_{n_1 n_{N+1}} = g(a_{n_i} - a_{n_{i+r+1}}) B_{n_1 n_2} B_{n_2 n_3} \cdots B_{n_N n_{N+1}} \,.
\]
This is a simple yet powerful result.
If the function \( g \) is replaced with \( \sinh \) or \( \coth \) functions, then \( \hat{G}_N \) may be determined without difficulty.
This allows calculations to be done numerically with relative ease.

%% file: conclusion.tex
\section{Conclusion}
A new representation for the Baker-Campbell-Hausdorff formula has been found.
This representation is a perturbative expansion in just one of two objects, as opposed to both in the original representation.
The series may then be truncated and give a good approximation to the full expansion for situations where only this second object is small.
For physical problems this then would give access to a much larger parameter space than is currently available.
Additionally, new problems for which the original representation was unusable may now be tackled.
Transfer matrices in statistical mechanics is an example of one such problem, which is under active consideration.

%% file: calculationOfTheSums.tex
\section{Calculation of the sums}\label{app:calculationOfTheSums}
Presented here is a direct method of calculating the sum \eqref{eq:explicitSum}.
As described in the main text, the key is to split the starting constrained sum into \( N + 1 \) semi-constrained sums (that is, one of the limits of the sum may be made infinite).
To that end, note
\[
  \sum_{-\frac{m-1}{2} \leq n_1 < n_2 < \cdots < n_N \leq \frac{m-1}{2}} = \sum_{n_1 < n_2 < \cdots < n_N \leq \frac{m-1}{2}} - \sum_{-\frac{m-1}{2} > n_1 < n_2 < \cdots < n_N \leq \frac{m-1}{2}} \,,
\]
which is demonstrated by the diagram below. In this, circles represent the variables of the sum and their position along the line indicates the value said variables take, while rectangles represent the bounds of the sums.
Open rectangles and circles allow equality, while filled do not.

\input{Identity1.tex}

This identity has transformed the constrained sum on the left into two sums.
One of these is semi-constrained, as was targeted, while the other has one semi-constrained and \( N-1 \) constrained variables.
Applying this idea again gives
\[
  \sum_{-\frac{m-1}{2} > n_1 < n_2 < \cdots < n_N \leq \frac{m-1}{2}} = 
    \sum_{\substack{  
      n_1 < -\frac{m-1}{2}\\ 
      n_2 < \cdots < n_N \leq \frac{m-1}{2}
    }} -
    \sum_{-\frac{m-1}{2} > n_1 \geq n_2 < \cdots < n_N \leq \frac{m-1}{2}} \,,
\]
or pictorially,

\input{Identity2.tex}

The number of constrained variables on the right hand side is now reduced to \( N-2\).
This can then be continued until there are no such variables remaining, resulting in an identity relating a sum with \( N \) constrained variables to \( N + 1 \) sums with only semi-constrained variables.

A generic term in this identity for the particular sum in the main text is given by
\begin{multline*}
   2^N 
  \sum_{\substack{  
    n_r \leq \cdots \leq n_1 < -\frac{m-1}{2}\\ 
    n_{r+1} < \cdots < n_N \leq \frac{m-1}{2}
  }}
  e^{2 n_1 L_1} e^{2 n_2 L_2} \cdots e^{2 n_N L_N}
  \\= 
  \left[ (-1)^r 2^{r} \sum_{n_r \leq \cdots \leq n_1 < -\frac{m-1}{2}} e^{2 n_1 L_1} \cdots e^{2 n_r L_r} \right]
  \left[ 2^{N-r} \sum_{n_{r+1} < \cdots < n_N \leq \frac{m-1}{2}} e^{2 n_{r+1} L_{r+1}} \cdots e^{2 n_N L_N} \right] \,,
\end{multline*}
that is,
\input{Identity3.tex}

A simple change of variables, indicated on the picture above, gives
\begin{multline*}
  \left[ (-1)^r 2^r \sum_{p_1=-\infty}^{-\frac{m-1}{2}-1} e^{2 p_1 ( L_1 + \cdots + L_r)} \sum_{p_2 = -\infty}^0 e^{2 p_2 ( L_2 + \cdots + L_r)} \cdots \sum_{p_r = -\infty}^0 e^{2 p_2 L_r} \right] \\
   \times \left[ 2^{N-r} \sum_{p_{r+1}=-\infty}^{-1} e^{2 p_{r+1} L_{r+1} } \cdots \sum_{p_{N-1} = -\infty}^{-1} e^{2 p_N-1 ( L_{r+1} + \cdots + L_{N-1})} \sum_{p_N = -\infty}^{\frac{m-1}{2}} e^{2 p_N (L_{r+1} + \cdots + L_{N})} \right] \,,
\end{multline*}
which may be trivially calculated.
Using the identities
\[
  -2 \sum_{n=-\infty}^{-\frac{m-1}{2} - 1} e^{2nx} =  \frac{-2 e^{-(m+1)x}}{1 - e^{-2x}} = \frac{e^{-mx}}{\sinh(-x)} \,, \qquad
  -2 \sum_{n=-\infty}^0 e^{2nx} = \frac{-2}{1 - e^{-2x}} = \coth(-x) - 1 \,,
\]
and
\[
  2 \sum_{n=-\infty}^{\frac{m-1}{2}} e^{2nx} =  \frac{2 e^{(m-1)x}}{1 - e^{-2x}} = \frac{e^{mx}}{\sinh(x)} \,, \qquad
  2 \sum_{n=-\infty}^{-1} e^{2nx} = \frac{2 e^{-2x}}{1 - e^{-2x}} = \coth(x) - 1 \,,
\]
provides the result required in the main text.

%% file: Identity1.tex
\begin{figure}[h]
\tikzset{
    ncbar angle/.initial=90,
    ncbar/.style={
        to path=(\tikztostart)
        -- ($(\tikztostart)!#1!\pgfkeysvalueof{/tikz/ncbar angle}:(\tikztotarget)$)
        -- ($(\tikztotarget)!($(\tikztostart)!#1!\pgfkeysvalueof{/tikz/ncbar angle}:(\tikztotarget)$)!\pgfkeysvalueof{/tikz/ncbar angle}:(\tikztostart)$)
        -- (\tikztotarget)
    },
    ncbar/.default=0.5cm,
}

\tikzset{square left brace/.style={ncbar=1cm}}
\tikzset{square right brace/.style={ncbar=-1cm}}
\tikzset{round left paren/.style={ncbar=.5cm,out=120,in=-120}}
\tikzset{round right paren/.style={ncbar=.5cm,out=60,in=-60}}
\centering
\begin{tikzpicture}[scale=0.25]
  \tikzstyle{every path}=[line width=.8pt]
  \def\height{12}

  \def\y{0}
  \def\x{0}
  \draw (\x,\height+\y) rectangle (\x+4,\height+1+\y); \node at (\x-1.5, \height+.5+\y) {\( \frac{m-1}{2} \)};
  \draw (\x+2,\height+\y) -- (\x+2,1+\y);
  \draw [fill=black] (\x+2,\height-1+\y) circle (.5); \node at (\x+4,\height-1+\y) {\( n_N \)};
  \draw [fill=black] (\x+2,\height-3+\y) circle (.5); \node at (\x+4.5,\height-3+\y) {\( n_{N-1} \)};
  \draw [fill=black] (\x+2,4+\y) circle (.5); \node at (\x+4,4+\y) {\( n_2 \)};
  \draw [fill=black] (\x+2,2+\y) circle (.5); \node at (\x+4,2+\y) {\( n_1 \)};
  \draw (\x,0+\y) rectangle (\x+4,1+\y); \node at (\x-2, .5+\y) {\( -\frac{m-1}{2} \)};
  
  \node at (8, \height/2+\y) {\( = \)};
  
  \def\x{12}
  \draw (\x,\height+\y) rectangle (\x+4,\height+1+\y); \node at (\x-1.5, \height+.5+\y) {\( \frac{m-1}{2} \)};
  \draw [->] (\x+2,\height+\y) -- (\x+2,0+\y);
  \draw [fill=black] (\x+2,\height-1+\y) circle (.5); \node at (\x+4,\height-1+\y) {\( n_N \)};
  \draw [fill=black] (\x+2,\height-3+\y) circle (.5); \node at (\x+4.5,\height-3+\y) {\( n_{N-1} \)};
  \draw [fill=black] (\x+2,4+\y) circle (.5); \node at (\x+4,4+\y) {\( n_2 \)};
  \draw [fill=black] (\x+2,2+\y) circle (.5); \node at (\x+4,2+\y) {\( n_1 \)};
  
  \node at (20, \height/2+\y) {\( - \)};

  \def\x{24}
  \draw (\x,\y-4.5) to [square left brace]  (\x,\y+\height+1.5);
  
  \def\x{28}
  \draw[dashed] (\x+2,-1+\y) -- (\x+10,-1+\y);
  \draw [fill=black] (\x,0+\y) rectangle (\x+4,1+\y); \node at (\x-2, .5+\y) {\( -\frac{m-1}{2} \)};
  \draw [->] (\x+2,0+\y) -- (\x+2,-4+\y);
  \draw [fill=white, draw=black] (\x+2,-1+\y) circle (.5); \node at (\x,-1+\y) {\( n_1 \)};
  
  \def\x{34}
  \draw (\x,\height+\y) rectangle (\x+4,\height+1+\y); \node at (\x-1.5, \height+.5+\y) {\( \frac{m-1}{2} \)};
  \draw (\x+2,\height+\y) -- (\x+2,-1+\y);
  \draw [fill=black] (\x+2,\height-1+\y) circle (.5); \node at (\x+4,\height-1+\y) {\( n_N \)};
  \draw [fill=black] (\x+2,\height-3+\y) circle (.5); \node at (\x+4.5,\height-3+\y) {\( n_{N-1} \)};
  \draw [fill=black] (\x+2,1+\y) circle (.5); \node at (\x+4,1+\y) {\( n_2 \)};
  
  \def\x{40}
  \draw (\x,\y-4.5) to [square right brace]  (\x,\y+\height+1.5);

\end{tikzpicture}
\end{figure}

%% file: Identity2.tex
\begin{figure}[h]
\tikzset{
    ncbar angle/.initial=90,
    ncbar/.style={
        to path=(\tikztostart)
        -- ($(\tikztostart)!#1!\pgfkeysvalueof{/tikz/ncbar angle}:(\tikztotarget)$)
        -- ($(\tikztotarget)!($(\tikztostart)!#1!\pgfkeysvalueof{/tikz/ncbar angle}:(\tikztotarget)$)!\pgfkeysvalueof{/tikz/ncbar angle}:(\tikztostart)$)
        -- (\tikztotarget)
    },
    ncbar/.default=0.5cm,
}

\tikzset{square left brace/.style={ncbar=1cm}}
\tikzset{square right brace/.style={ncbar=-1cm}}
\tikzset{round left paren/.style={ncbar=.5cm,out=120,in=-120}}
\tikzset{round right paren/.style={ncbar=.5cm,out=60,in=-60}}
\centering
\begin{tikzpicture}[scale=0.25]
  \tikzstyle{every path}=[line width=.8pt]
  \def\height{12}
  
  \def\y{0}
  \def\x{0}
  \draw (\x,\y-4.5) to [square left brace]  (\x,\y+\height+1.5);
  
  \def\x{4}
  \draw [fill=black] (\x,0+\y) rectangle (\x+4,1+\y); \node at (\x-2, .5+\y) {\( -\frac{m-1}{2} \)};
  \draw [->] (\x+2,0+\y) -- (\x+2,-4+\y);
  \draw [fill=white, draw=black] (\x+2,-1+\y) circle (.5); \node at (\x+4,-1+\y) {\( n_1 \)};
  
  \def\x{10}
  \draw (\x,\height+\y) rectangle (\x+4,\height+1+\y); \node at (\x-1.5, \height+.5+\y) {\( \frac{m-1}{2} \)};
  \draw (\x+2,\height+\y) -- (\x+2,-1+\y);
  \draw [fill=black] (\x+2,\height-1+\y) circle (.5); \node at (\x+4,\height-1+\y) {\( n_N \)};
  \draw [fill=black] (\x+2,\height-3+\y) circle (.5); \node at (\x+4.5,\height-3+\y) {\( n_{N-1} \)};
  \draw [fill=black] (\x+2,1+\y) circle (.5); \node at (\x+4,1+\y) {\( n_2 \)};
  \draw [dashed] (\x-1,-1+\y) -- (\x+2,-1+\y);
  
  \def\x{16}
  \draw (\x,\y-4.5) to [square right brace]  (\x,\y+\height+1.5);
  
  \def\x{20}
  \node at (\x, \height/2+\y) {\( = \)};

  \def\x{24}
  \draw (\x,\y-4.5) to [square left brace]  (\x,\y+\height+1.5);
  
  \def\x{28}
  \draw [fill=black] (\x,0+\y) rectangle (\x+4,1+\y); \node at (\x-2, .5+\y) {\( -\frac{m-1}{2} \)};
  \draw [->] (\x+2,0+\y) -- (\x+2,-4+\y);
  \draw [fill=white, draw=black] (\x+2,-1+\y) circle (.5); \node at (\x+4,-1+\y) {\( n_1 \)};
  
  \def\x{34}
  \draw (\x,\height+\y) rectangle (\x+4,\height+1+\y); \node at (\x-1.5, \height+.5+\y) {\( \frac{m-1}{2} \)};
  \draw [->] (\x+2,\height+\y) -- (\x+2,-4+\y);
  \draw [fill=black] (\x+2,\height-1+\y) circle (.5); \node at (\x+4,\height-1+\y) {\( n_N \)};
  \draw [fill=black] (\x+2,\height-3+\y) circle (.5); \node at (\x+4.5,\height-3+\y) {\( n_{N-1} \)};
  \draw [fill=black] (\x+2,-2+\y) circle (.5); \node at (\x+4,-2+\y) {\( n_2 \)};
  
  \def\x{40}
  \draw (\x,\y-4.5) to [square right brace]  (\x,\y+\height+1.5);
  
  \def\x{44}
  \node at (\x, \height/2+\y) {\( - \)};
  
  \def\x{48}
  \draw (\x,\y-4.5) to [square left brace]  (\x,\y+\height+1.5);
  
  \def\x{52}
  \draw [fill=black] (\x,0+\y) rectangle (\x+4,1+\y); \node at (\x-2, .5+\y) {\( -\frac{m-1}{2} \)};
  \draw [->] (\x+2,0+\y) -- (\x+2,-6+\y);
  \draw [fill=white, draw=black] (\x+2,-1+\y) circle (.5); \node at (\x+4,-1+\y) {\( n_1 \)};
  \draw [fill=white, draw=black] (\x+2,-3+\y) circle (.5); \node at (\x+4,-3+\y) {\( n_2 \)};
  
  \def\x{58}
  \draw (\x,\height+\y) rectangle (\x+4,\height+1+\y); \node at (\x-1.5, \height+.5+\y) {\( \frac{m-1}{2} \)};
  \draw (\x+2,\height+\y) -- (\x+2,-3+\y);
  \draw [fill=black] (\x+2,\height-1+\y) circle (.5); \node at (\x+4,\height-1+\y) {\( n_N \)};
  \draw [fill=black] (\x+2,\height-3+\y) circle (.5); \node at (\x+4.5,\height-3+\y) {\( n_{N-1} \)};
  \draw [fill=black] (\x+2,-1+\y) circle (.5); \node at (\x+4,-1+\y) {\( n_3 \)};
  \draw [dashed] (\x,-3+\y) rectangle (\x+4,-3+\y);

  \def\x{64}
  \draw (\x,\y-4.5) to [square right brace]  (\x,\y+\height+1.5);

\end{tikzpicture}
\end{figure}

%% file: Identity3.tex
\begin{figure}[h]
\tikzset{
    ncbar angle/.initial=90,
    ncbar/.style={
        to path=(\tikztostart)
        -- ($(\tikztostart)!#1!\pgfkeysvalueof{/tikz/ncbar angle}:(\tikztotarget)$)
        -- ($(\tikztotarget)!($(\tikztostart)!#1!\pgfkeysvalueof{/tikz/ncbar angle}:(\tikztotarget)$)!\pgfkeysvalueof{/tikz/ncbar angle}:(\tikztostart)$)
        -- (\tikztotarget)
    },
    ncbar/.default=0.5cm,
}

\tikzset{square left brace/.style={ncbar=1cm}}
\tikzset{square right brace/.style={ncbar=-1cm}}
\tikzset{round left paren/.style={ncbar=.5cm,out=120,in=-120}}
\tikzset{round right paren/.style={ncbar=.5cm,out=60,in=-60}}
\centering
\begin{tikzpicture}[scale=0.25]
  \tikzstyle{every path}=[line width=.8pt]
  \def\height{12}
  
  \def\y{0}
  \def\x{0}
  \draw (\x,\y-11.5) to [square left brace]  (\x,\y+\height+1.5);
  
  \def\x{4}
  \draw [fill=black] (\x,0+\y) rectangle (\x+4,1+\y); \node at (\x-2, .5+\y) {\( -\frac{m-1}{2} \)};
  \draw [->] (\x+2,0+\y) -- (\x+2,-11+\y);
  \draw [fill=white, draw=black] (\x+2,-1.5+\y) circle (.5); \node at (\x+4,-1.5+\y) {\( n_1 \)};
  \draw [fill=white, draw=black] (\x+2,-3.5+\y) circle (.5); \node at (\x+4,-3.5+\y) {\( n_2 \)};
  \draw [fill=white, draw=black] (\x+2,-7+\y) circle (.5); \node at (\x+4,-7+\y) {\( n_{r-1} \)};
  \draw [fill=white, draw=black] (\x+2,-9+\y) circle (.5); \node at (\x+4,-9+\y) {\( n_r \)};
  \draw [<->] (\x+6, .5\y) -- (\x+6, -1.5+\y); \node at (\x+7, -.5+\y) {\( p_1 \)};
  \draw [<->] (\x+6, -1.5+\y) -- (\x+6, -3.5+\y); \node at (\x+7, -2.5+\y) {\( p_2 \)};
  \draw [<->] (\x+6, -7+\y) -- (\x+6, -9+\y); \node at (\x+7, -8+\y) {\( p_r \)};
  
  \def\x{12}
  \draw (\x,\height+\y) rectangle (\x+4,\height+1+\y); \node at (\x-1.5, \height+.5+\y) {\( \frac{m-1}{2} \)};
  \draw [->] (\x+2,\height+\y) -- (\x+2,-11+\y);
  \draw [fill=black] (\x+2,\height-1.5+\y) circle (.5); \node at (\x+4,\height-1.5+\y) {\( n_N \)};
  \draw [fill=black] (\x+2,\height-3.5+\y) circle (.5); \node at (\x+4.5,\height-3.5+\y) {\( n_{N-1} \)};
  \draw [fill=black] (\x+2,-7+\y) circle (.5); \node at (\x+4,-7+\y) {\( n_{r+2} \)};
  \draw [fill=black] (\x+2,-9+\y) circle (.5); \node at (\x+4,-9+\y) {\( n_{r+1} \)};
  \draw [<->] (\x+6, \height+.5+\y) -- (\x+6,\height-1.5+\y); \node at (\x+7, \height-.5+\y) {\( p_N \)};
  \draw [<->] (\x+6, \height-1.5+\y) -- (\x+6, \height-3.5+\y); \node at (\x+8, \height-2.5+\y) {\( p_{N-1} \)};
  \draw [<->] (\x+6, -7+\y) -- (\x+6, -9+\y); \node at (\x+8, -8+\y) {\( p_{r+1} \)};
  
  \def\x{21}
  \draw (\x,\y-11.5) to [square right brace]  (\x,\y+\height+1.5);

\end{tikzpicture}
\end{figure}